\documentclass[floatfix,amsfonts,amsmath,preprintnumbers,nofootinbib,letterpaper,superscriptaddress,11pt]{revtex4}
\usepackage{amsmath,amssymb}
\usepackage{graphicx}
\usepackage{bm}
\usepackage{comment}
\usepackage{color}

\newcommand{\beq}{\begin{equation}}
\newcommand{\eeq}{\end{equation}}
\newcommand{\bea}{\begin{eqnarray}}
\newcommand{\eea}{\end{eqnarray}}

\def\OMIT#1{{}}

\begin{document}

\preprint{NT@UW-08-18, DAMTP-2008-94}

\title{Bottom hadron mass splittings in the static limit
  from 2+1 flavour lattice QCD}
\author{William Detmold}
\affiliation{Department of Physics, University of Washington, Seattle,
  WA 98195-1560, USA}
\affiliation{Department of Physics, College of William \&  Mary, Williamsburg,
  VA 23187-8795, USA}
\author{C.-J. David Lin}
\affiliation{Institute of Physics, National Chiao-Tung University, 
Hsinchu 300, Taiwan}
\affiliation{Physics Division, National Centre for Theoretical Sciences,
Hsinchu 300, Taiwan}
\author{Matthew Wingate}
\affiliation{DAMTP, University of Cambridge,
Wilberforce Road, Cambridge CB3 0WA, UK}
\date{\today}

\begin{abstract}
\noindent 
Dynamical 2+1 flavour lattice QCD is used to calculate the splittings
between the masses of mesons and baryons containing a single static
heavy quark and domain-wall light and strange quarks. Our calculations
are based on the dynamical domain-wall gauge field configurations
generated by the RBC and UKQCD collaborations at a spatial volume of
(2.7 fm)$^3$ and a range of quark masses with a lightest value
corresponding to a (partially-quenched) pion mass of $275$~MeV. When
extrapolated to the physical values of the light quark masses, the
results of our 
calculations are generally in good agreement with experimental
determinations in the bottom sector. However, the static limit
splittings between the $\Omega_b^-$ baryon and other bottom hadrons
tend to slightly underestimate those obtained using the recent
D$0\!\!\!/$ measurement of the $\Omega_b^-$.
\end{abstract}
\maketitle

%
%

\section{Introduction}
\label{sec:introduction}

Hadrons containing a single bottom quark have received much attention
recently. Over the last decade, the bottom meson sector has been
investigated in great detail at the $b$ factories (BELLE and BaBar)
and using the TeVatron at Fermilab.  With the recent D$0\!\!\!/$
measurement of the $\Omega_b^-$ baryon \cite{Abazov:2008qm}, we are
rapidly approaching a complete picture of the ground-state bottom
baryon sector as well. The start up of the Large Hadron Collider (LHC)
will dramatically increase our knowledge of $b$ physics; both the
dedicated $b$ physics experiment, LHCb, and the general purpose ATLAS
and CMS experiments expect to observe unprecedented numbers of
$b\overline{b}$ pairs, many of which will hadronise to bottom mesons
and baryons.

In the bottom meson sector, many important observations have been
made in the last decade, dramatically refining our knowledge of
flavour physics.  The CKM mechanism of the Standard Model currently
provides a good description of current flavour-changing measurements
of $B$ and $B_s$ mesons, bringing us into the
precision era of flavour physics \cite{Barberio:2008fa}. The various
LHC experiments aim to investigate the decays of bottom baryons with
enough precision to further test the Standard Model. In particular,
the fact that bottom baryon polarisation 
can be measured may uncover new right-handed couplings.  To search for
physics beyond the Standard Model in these systems, it is necessary to
have accurate predictions from the Standard Model. Typically, this
requires non-perturbative evaluations of matrix elements from lattice
QCD and thus an understanding of bottom hadrons in lattice
QCD.

The simplest properties of the bottom hadrons are their masses.
Before more complicated bottom observables can be predicted with any
rigour, it is necessary to have reliable lattice calculations of the
masses and mass splittings. There are many lattice studies of the
mass of the $B$ mesons in
the literature but fewer of the bottom baryons \cite{Bowler:1996ws,AliKhan:1999yb,
  Lewis:2001iz,Mathur:2002ce} and, only recently, 
the first unquenched bottom baryon calculations have appeared
\cite{Na:2007pv,Lewis:2008fu,Burch:2008qx,Na:2008hz}. In
this paper, we use light up and down quark masses to study the
spectrum of bottom baryons and mesons. For the bottom quark, we work in the 
static ($m_b\to\infty$) limit. Since the bottom
quark mass, $m_b\gg \Lambda_{\mathrm{QCD}}$, many properties of the physical
bottom hadrons are expected to be close to those of hadrons containing
a single, infinitely massive (static) quark, with corrections
suppressed by powers of $\Lambda_{\mathrm{QCD}}/m_b$. Such corrections can be
addressed systematically in Heavy Quark Effective Theory (see, {\it
  e.g.} Ref.~\cite{Manohar:2000dt} for a review). Our calculations use
domain-wall fermions  for the light and strange quarks and
make use of the dynamical gauge configurations generated by the RBC
and UKQCD collaborations using domain-wall quarks \cite{Antonio:2006px,RBCUKQCD}. We
currently work at a single lattice spacing, $a=0.114$~fm
\cite{RBCUKQCD} and focus on a volume of spatial side, $L=2.74$~fm,
for a range of quark masses corresponding to pion masses between 275
and 750~MeV.

This paper is organised in the following manner. In Section
\ref{sec:deta-latt-calc}, we present the details of the lattice
calculations we have carried out. We discuss the mass splittings we
observe in Section~\ref{sec:su3-mass-splittings} before concluding in
Section~\ref{sec:conclusions-outlook}.

\section{Details of Lattice Calculations}
\label{sec:deta-latt-calc}

The lattice QCD calculations presented here are based on the ensembles
of 2+1 flavour lattices generated by the RBC and UKQCD collaborations
\cite{Antonio:2006px,RBCUKQCD}. These ensembles use the Iwasaki gauge action
\cite{Iwasaki:1984cj,Iwasaki:1985we} with $\beta=2.13$ and a domain
wall quark action using the length of the fifth dimension,
$L_s=16$. The parameters of the ensembles used this work are shown in
Table \ref{tab:ensemble} and we refer the reader to
Ref.~\cite{Antonio:2006px,RBCUKQCD} for further details.
\begin{table}[h]
  \begin{ruledtabular}
  \begin{tabular}{ccccccc}
    Ensemble & $\beta$ & Volume  & $m_{\rm light}^{(sea)}$ &
    $m_{\rm strange}^{(sea)}$ & $m_{\rm res}$ \cite{RBCUKQCD} & $N_{\rm cfgs}$ \\
\hline
A & 2.13  & $24^3\times64$ & 0.005 & 0.04 & 0.00315  &
$\sim 140$ \\
B & 2.13  & $24^3\times64$ & 0.01 & 0.04 & 0.00315 &
$\sim 180$ \\
C & 2.13  & $24^3\times64$ & 0.02 & 0.04 & 0.00320  &
$\sim 125$ \\
  \end{tabular}
  \end{ruledtabular}
  \caption{Parameters of the ensembles of RBC/UKQCD gauge configurations used in
    this calculation.  For full details, see the
    original works, Refs.~\protect{\cite{Antonio:2006px,RBCUKQCD}}.}
  \label{tab:ensemble}
\end{table}

Using these ensembles, we have computed domain wall quark propagators
\cite{Kaplan:1992bt,Shamir:1992im,Shamir:1993zy,Shamir:1998ww,Furman:1994ky}
for various different (partially quenched) quark masses as shown in
Table \ref{tab:props}. Our inversions use $L_s=16$ and a domain wall
height of $M_{5}=1.8$ (note that these are the same parameters used in
generating the ensembles and the bold entries in Table \ref{tab:props}
correspond to QCD computations). To obtain clean signals, for the
hadron energies and splittings, we use APE
\cite{Teper:1987wt,Albanese:1987ds} smeared sources at multiple
locations on each gauge configuration, performing a separate inversion
for each source. The approximate masses of the pseudoscalar mesons computed with
various combinations of these propagators are shown in
Table~\ref{tab:piK}.
\begin{table}[h]
  \centering
  \begin{ruledtabular}
  \begin{tabular}{ccccc}
    Ensemble & $m_{\rm light}^{(val)}$ 
       &  $N_{\rm cfg}$ & $N_{\rm src}$ \\ \hline
        A &   0.002 &   $\sim140$ & 5 \\
        {\bf A} &   {\bf 0.005} &   $\sim$ {\bf 140} & {\bf 6} \\
        A &   0.01  &   $\sim140$ & 1 \\
        A &   0.02  &   $\sim140$ & 1 \\
        A &   0.03  &   $\sim140$ & 1 \\
        {\bf A} &   {\bf 0.04}  &   $\sim${\bf 140} & {\bf 6} \\ \hline
        B &   0.005   &   $\sim180$ & 1 \\
        {\bf B} &   {\bf 0.01}   &   $\sim${\bf 180} & {\bf 5} \\
        B &   0.02   &   $\sim180$ & 1 \\
        B &   0.03   &   $\sim180$ & 1 \\
        {\bf B} &   {\bf 0.04}   &   $\sim${\bf 180} & {\bf 5}     \\ \hline
        {\bf C} &   {\bf 0.02}   &   $\sim${\bf 125} & {\bf 1} \\
        {\bf C} &   {\bf 0.04}   &   $\sim${\bf 125} & {\bf 1} 
  \end{tabular}
  \end{ruledtabular}
  \caption{Quark propagators used in calculations of the mass
    splittings. Bold entries denote $m_{\rm light}^{(val)} = 
    m_{\rm light}^{(sea)}$.  $N_{\rm src}$ indicates the number of sources that
    were used on each gauge configuration. These sources were spread
    around the lattice volume at as large relative separations as possible.}
  \label{tab:props}
\end{table}

\begin{table}
  \centering
  \begin{ruledtabular}
  \begin{tabular}{ccccc}
    Ensemble & $m^{(\rm val)}_{\rm light}$ & $m^{(\rm val)}_{\rm
      strange}$  & $m^{(\rm val)}_\pi$ [GeV] & $m^{(\rm val)}_K$ [GeV] \\ \hline
    A & 0.002 & 0.04 & 0.275 & 0.560 \\
    A & 0.005 & 0.04 & 0.331 & 0.576 \\
    A & 0.01  & 0.04 & 0.415 & 0.602 \\
    A & 0.02  & 0.04 & 0.546 & 0.654 \\
    A & 0.03  & 0.04 & 0.653 & 0.701 \\
    A & 0.04  & 0.04 & 0.747 & ---\\ \hline
    B & 0.005 & 0.04 & 0.335 & 0.581 \\
    B & 0.01  & 0.04 & 0.419 & 0.607 \\
    B & 0.02  & 0.04 & 0.550 & 0.657 \\
    B & 0.03  & 0.04 & 0.657 & 0.706 \\
    B & 0.04  & 0.04 & 0.751 & --- \\ \hline
    C & 0.02  & 0.04 & 0.549 & 0.654 
  \end{tabular}
  \end{ruledtabular}
  \caption{Approximate pion and kaon masses for each set of measurements.}
  \label{tab:piK}
\end{table}

The bottom quark is implemented in the static limit. Its propagator,
$S_Q$ is represented as a product of gauge links in the temporal
direction,
\begin{equation}
  \label{eq:4}
  S_Q({\bf x},t;t_0)=\left(\frac{1+\gamma_4}{2}\right)\prod_{t^\prime=t_0}^{t} 
U_4({\bf x},t^\prime)\,,
\end{equation}
where $U_\mu$ are SU(3) gauge links.  At non-zero lattice spacing,
there are different discretisations of the heavy quark
action that have the same continuum limit. It is well-known (see {\it
  e.g.} Ref.~\cite{DellaMorte:2003mn}) that signals for static hadron
quantities are improved if the gauge links appearing in
Eq.~(\ref{eq:4}) are smeared in some manner over a small local
volume. After extensive testing to optimise the heavy hadron signals,
we perform our calculations with hypercubically-smeared (HYP) gauge
links \cite{Hasenfratz:2001hp} in the heavy quark propagator,
Eq.~(\ref{eq:4}).  We study a number of choices of heavy quark
smearing parameters as shown in Table \ref{tab:smear}, labelled $S_i$
for $i=0,\ldots,5$.
\begin{table}[h]
  \centering
  \begin{ruledtabular}
  \begin{tabular}{c||ccccccc}
    Set & $S_0$ & $S_1$ & $S_2$ & $S_3$ & $S_4$ & $S_5$\\
    \hline
    $N_{\rm HYP}$ & 10 & 5 & 1 & 0 & 3 & 3
    \\
    $\alpha_1$ & 0.75 & 0.75 & 0.75 & 0.75 & 0.75 & 0.6
    \\
    $\alpha_2$ & 0.75 & 0.75 & 0.75 & 0.75 & 0.4 & 0.6
    \\
    $\alpha_3$ & 0.75 & 0.75 & 0.75 & 0.75 & 0.65 & 0.6
  \end{tabular}
  \end{ruledtabular}
  \caption{Parameters used in HYP-smearing the heavy quark action.}
  \label{tab:smear}
\end{table}

To extract the lattice energies, $E_h$, of the various hadrons, $h$,
we compute the meson and baryon two-point correlation functions
\begin{eqnarray}
  \label{eq:5}
  C_f(t,t_0) &=& \sum_{{\bf x}} {\rm tr}\left[S_Q({\bf x},t;t_0)
   S_f^\dagger({\bf x},t;{\bf x},t_0) \right]\,,
\\
\label{eq:6}
  C^\Gamma_{f,g}(t,t_0) &=& \sum_{{\bf x}} 
    S^{k^\prime k}_{Q;\sigma\rho}({\bf x},t;t_0)\epsilon^{ijk}
    \epsilon^{i^\prime j^\prime k^\prime}
    \left(S_f^{i i^\prime}({\bf x},t;{\bf x},t_0) \Gamma\right)_{\rho\alpha} 
    \left(\Gamma S_g^{j j^\prime}({\bf x},t;{\bf x},t_0)\right)_{\sigma\alpha}  
\, ,
\end{eqnarray}
for the various light flavour combinations, $f$($g$), and then combine
them appropriately.  The trace in the meson correlator is over color
and spinor indices, and where explicit, the upper(additional lower)
indices on the propagators correspond to colour(spin). For the baryons
with the spin of the light degrees of freedom, $s_\ell$, being zero we
choose $\Gamma=C\, \gamma_5$ where $C$ is the charge conjugation
matrix, while for the baryons with $s_\ell=1$, we measure the three
polarisations, corresponding to $\Gamma=C\,\gamma_{1,2,3}$, and
average them in extracting the energies as they are degenerate in the
infinite statistics limit. We combine the various correlators in
Eq.~(\ref{eq:6}) as appropriate for the specific
SU(3) representation. For example, the $\Xi_b^{0}$ belongs to the flavour
anti-triplet and is a $s_\ell=0$ state, consequently
\begin{equation}
  \label{eq:1}
  C_{\Xi_b^{ 0}}= C^{\gamma_5}_{u,s} -C^{\gamma_5}_{s,u}
\end{equation}
and similarly,
\begin{eqnarray}
  \label{eq:2}
    C_{\Lambda_b^{ 0}}= C^{\gamma_5}_{u,d} -C^{\gamma_5}_{d,u}, \quad \quad
    C_{\Sigma_{b,i}^{0}}= C^{\gamma_i}_{u,d} +C^{\gamma_i}_{d,u}, \quad\quad
    C_{\Xi_{b,i}^{\prime,0}}= C^{\gamma_i}_{u,s} +C^{\gamma_i}_{s,u}, \quad\quad
    C_{\Omega_{b,i}^{-}}= C^{\gamma_i}_{s,s} \,.
\end{eqnarray}
As we work in the isospin limit, the other singly-heavy hadrons are
degenerate with those above. In the static limit of the heavy quark,
the $J^P=\frac{1}{2}^+$ states ($\Sigma_b,\,\Xi_b^\prime,\, \Omega_b$)
and $\frac{3}{2}^+$ states ($\Sigma_b^\ast,\,\Xi_b^{\prime\ast},\,
\Omega_b^\ast$) are degenerate and our results correspond to
this. Future calculations for $m_b\ne\infty$ will enable the spin
splittings to be investigated, see Ref.~\cite{Lewis:2008fu} for recent
work.

The transfer matrix formalism immediately shows that in the limit of
large temporal separation, these correlators are dominated by
exponentially decaying signal of the ground-state hadron energy. That
is,
\begin{equation}
  \label{eq:3}
  C_h(t,t_0)\stackrel{t\gg t_0}{\longrightarrow} A_h \exp{[-E_h(t-t_0)]}\,,
\end{equation}
where $E_h$ is the lattice energy. In the case of the $s_\ell=1$
baryons of strangeness $<2$, this is complicated by the fact that, at
the physical quark masses and in infinite volume, they decay strongly
to lighter $s_\ell=0,1$ baryons and a pion. For example,
$\Sigma_b^+\to\Lambda_b\, \pi^+$ is the dominant decay mode
\cite{Amsler:2008zz}. However, at the unphysical masses and finite
volumes used (the pion couples derivatively, requiring non-zero
lattice momentum) here, these decays are forbidden. Additionally, the
finite temporal extent (non-zero temperature) of the lattice geometry
used in our calculations results in the appearance of exponentially suppressed states
that decay with ``lighter masses'' that may pollute the signals of the
zero temperature ground-state \cite{Detmold:2008aaa} (this is
asymmetric analog of the constant term expected in a two meson
correlation function from one meson propagating forward in time and
one propagating backward in time
\cite{Detmold:2008yn,Prelovsek:2008rf}).  We will return to this issue
below.

\section{Bottom Hadron Mass Splittings}
\label{sec:su3-mass-splittings}

In the static limit, the masses of bottom hadrons are infinite,
however, the residual mass, $\overline{\Lambda}_h$, of a heavy hadron
$h$ is well defined. In the lattice regularisation used here, the
lattice energies, $E_h$ diverge as the continuum limit is taken and a
perturbative subtraction is required to convert them to the
corresponding residual mass, for example, in the modified minimal
subtraction ($\overline{\rm MS}$) scheme. At one loop, this can be
performed straightforwardly for typical lattice actions, and with more
effort for the HYP-smeared gauge links used herein
\cite{Loktik:2006kz,Christ:2007cn}. However, there are significant
issues with higher order corrections and renormalons in the
subtraction procedure \cite{Martinelli:1998vt, Martinelli:1995vj}. 
In contrast, the differences
between the residual masses of two hadrons containing the same heavy
quark (that is, the same lattice discretisation of a static quark)
have well-defined continuum limits in which they are observable (the
perturbative corrections and renormalisation scale dependence cancel
in the difference). In our work, we concentrate on these differences
and thereby avoid the issues mentioned above.

\subsection{Heavy quark action}
\label{sec:heavy-quark-action}

\begin{figure}
  \centering
  \includegraphics[width=\columnwidth]{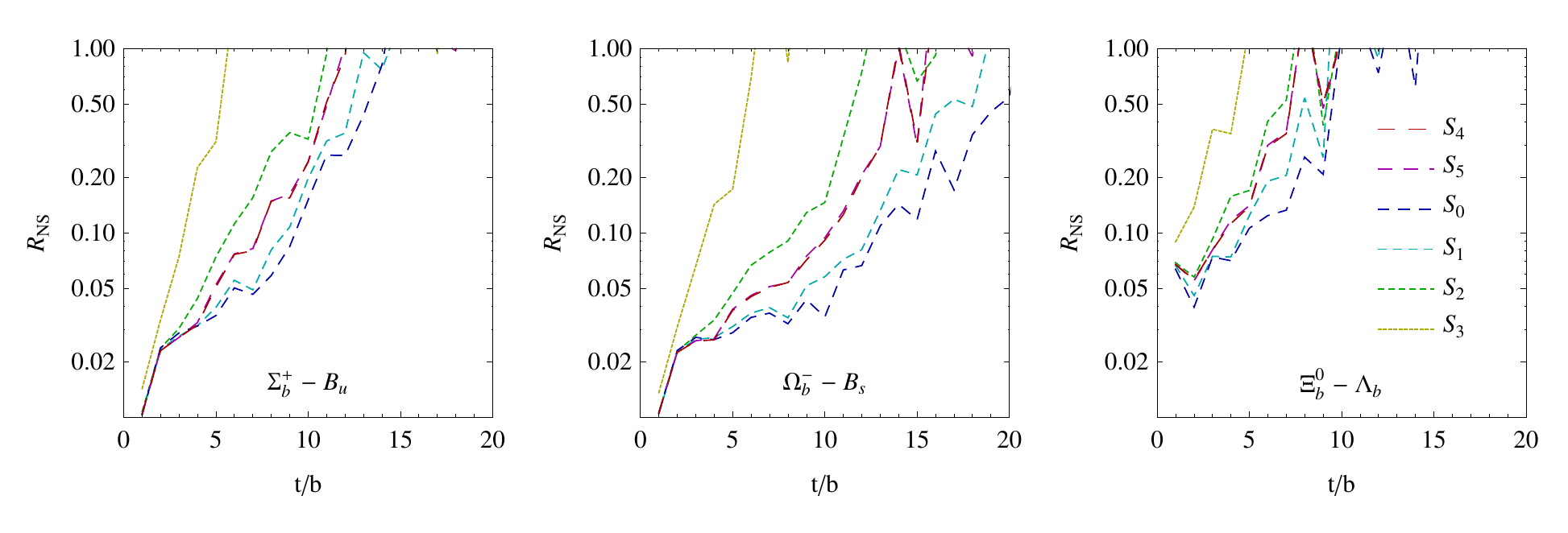}
  \caption{The ratio of noise to signal, $R_{NS}$, for exemplary
    correlator differences for the different choices of the heavy
    quark action (Table~\protect{\ref{tab:smear}}). The $m^{(\rm
      val)}=m^{(\rm sea)}$ quark propagators on the B ensemble were
    used in this comparison. The other sets display similar
    behaviour.}
\label{fig:s2n}
\end{figure}

We first study the effects of the choice of the heavy quark action via
the application of the various HYP-smearings shown in
Table~\ref{tab:smear}. For each energy or energy difference, the
results for all the choices of smearing are found to be consistent within
uncertainties. Figure~\ref{fig:s2n} shows the noise-to-signal ratios
for the correlators corresponding to the $\Sigma_b^+$---$B_u$,
$\Omega_b^-$---$B_s$ and $\Xi_b^0$---$\Lambda_b$ mass splittings for
the various smearing parameter sets as a function of Euclidean time
using the QCD propagators of ensemble B in Table
\ref{tab:ensemble}. As can be seen, the parameter set $S_0$ clearly
has the strongest signals. The other sets of propagators display
similar behaviour and we focus on set $S_0$ for the remainder of our
study.

\subsection{Analysis Methods}
\label{sec:analysis-methods}

To extract the mass differences from the lattice correlators defined
in the previous section, we perform multiple independent analyses
using different methods. We fit to either single- or two-state
effective masses and /or mass differences(see Ref.~\cite{Fleming:2004hs}), or perform
multi-exponential fits to (differences of) correlation functions using
Bayesian priors (as in Refs~\cite{Lepage:2001ym,Wingate:2002fh}).  We
use the bootstrap analysis method to estimate our uncertainties. We
average over all propagators computed from the various sources on each
configuration and then bin the measurements over units of $\sim$100
molecular dynamics time units as residual correlations are observed to
persist over this range \cite{RBCUKQCD}.  The number of bootstrap
ensembles used in our analysis is typically four times the number of binned
measurements, $N_{\rm bootstrap}\sim200$.

Correlation functions are calculated with both an APE smeared sink
(using the same parameters as at the source) or a point sink. Results
from both sink types are consistent. In general, the fitting to the
energy differences leads to a cleaner signal than fitting each energy
individually, but not in all cases and both approaches were tried.

In the correlated fits to single- and two-state effective masses, a
systematic uncertainty 
is assigned from varying the fitting ranges, $t_{\rm min}$ to $t_{\rm
  max}$, or from differences between fits to sliding windows of
time-slices within the overall fit range.  Typically, single effective
masses are fit from $t_{\rm min}\sim9$ and two-state effective masses
are fit from $t_{\rm min}\sim4$ (here, the ground state
plateaus earlier as the excited state contamination is decreased
significantly). The upper limit of the fits is $t_{\rm max}\sim20$ for
mesons and anti-triplet baryons, and $t_{\rm max}\sim15$ for sextet
baryons where states arising from the finite temporal extent discussed
above pollute the signal.

In the Bayesian analysis, the correlation functions are fit to
\begin{equation}
C(t) ~=~ A_1 e^{-E_1 t}(1 + B_1 e^{-\Delta E_1 t} 
+ B_2 e^{-(\Delta E_1 +\Delta E_2) t}) + A_1C_1e^{-E' t} \,.
\label{eq:fitfunction}
\end{equation}
over the range $2\le t \le 20$. This form was sufficient to provide an
acceptable fit to all correlators and additional exponential terms
were not constrained by the data. The Bayesian prior distribution
functions are Gaussian.  The mean values for the $A_1$ and $E_1$
priors were estimated from single exponential fits at large times.
The widths for those parameters were taken to be about 50\% and 10\%,
respectively, are significantly larger than the
uncertainties from the initial single exponential fit.  The mean
values for the $B_i$ priors were in $[-1,1]$ with widths up to a
factor 10, and the $\Delta E_i$ priors had means in $[0.3,0.6]$ with
widths between 60\% and 80\%. 
The last term in (\ref{eq:fitfunction}) allows for contributions to
$C(t)$ due to multi-particle states created by the interpolating
operators in Eq~(\ref{eq:5}) and (\ref{eq:6}).  Usually these states
give vanishingly small contribution, 
as their total energy is greater than the hadron mass of interest.
However, with (anti-)periodic boundary conditions in $t$, these states can
contaminate the large $t$ behavior due to a backward propagating light
meson \cite{Detmold:2008aaa}.  Therefore we include the last term in
Eq.~(\ref{eq:fitfunction}) with prior mean for $C_1 \approx
\exp(-0.1\, T)$, where $T$ is the temporal extent of the lattice,
and for $E'=0.5$, both with 100\% widths.

In combining the various analyses, we make a direct comparison of the
various extractions and if
they are in agreement, we take their mean. If there are discrepancies
(very few are found), we assign an additional systematic uncertainty
large enough to make the result consistent with the individual
analyses.

The final results of our analyses of the lattice calculations are
shown in Fig.~\ref{fig:massdep} for the various quark masses on the
three ensembles (the statistical, and all systematic uncertainties have been
combined in quadrature). The empty points correspond to calculations
that are partially quenched, with the sea quark masses differing from
the valence quark masses. Ensembles A, B and C are coloured blue, red
and green respectively.

\subsection{Quark mass extrapolations}
\label{sec:quark-mass-extr}

The light quark mass dependence of the static hadron energies and mass
differences calculated in this work is, at least in principle,
described by the low energy effective theory, heavy hadron chiral
perturbation theory (HH$\chi$PT)
\cite{Burdman:1992gh,Wise:1992hn,Yan:1992gz,Cho:1992gg,Cho:1992cf} or
its quenched and partially-quenched versions
\cite{Savage:1995dw,Chiladze:1997uq,Tiburzi:2004kd}. In addition to
polynomial dependence on the pion and kaon masses, non-analytic terms
such as $m_\pi^2\log[m_\pi^2/\mu^2]$ appear ($\mu$ is the
renormalisation scale). We have attempted to perform extrapolations
using the forms predicted by these theories. Unfortunately, our
current data are insufficient to constrain the coefficients of the
non-analytic terms in the HH$\chi$PT expressions (three axial
couplings contribute in the coefficients of the non-analytic terms in
the various differences). Future, direct calculations of the axial
couplings will enable a more controlled chiral extrapolation.

With this in mind, here we use simple polynomial fits to perform the
light quark mass extrapolations, allowing linear, quadratic and cubic
dependence on $m_q\sim m_\pi^2$. Since we keep $m_s$ fixed, we do not
include dependence on this parameter (we note that it is not quite
tuned to the physical value \cite{RBCUKQCD}). These fits use a
maximum likelihood estimator and are performed in an uncorrelated
manner for simplicity. Accounting for the correlations between the
different partially-quenched data points computed on the
same underlying ensemble would slightly reduce our
uncertainty. However, given the extrapolation forms used at present
are {\it ad hoc}, we do not pursue this further. The resulting
$\chi^2$s of the fits are all acceptable.

We have also performed coupled fits to all the differences, allowing a
fixed polynomial dependence on $m_\pi$ for each of the hadron
energies. This coupled fitting procedure provides a successful fit
and agrees with the above (uncoupled) method within uncertainties but
does not improve the extrapolations.

\begin{figure}[!t]
  \centering
    \includegraphics[width=0.80\columnwidth]{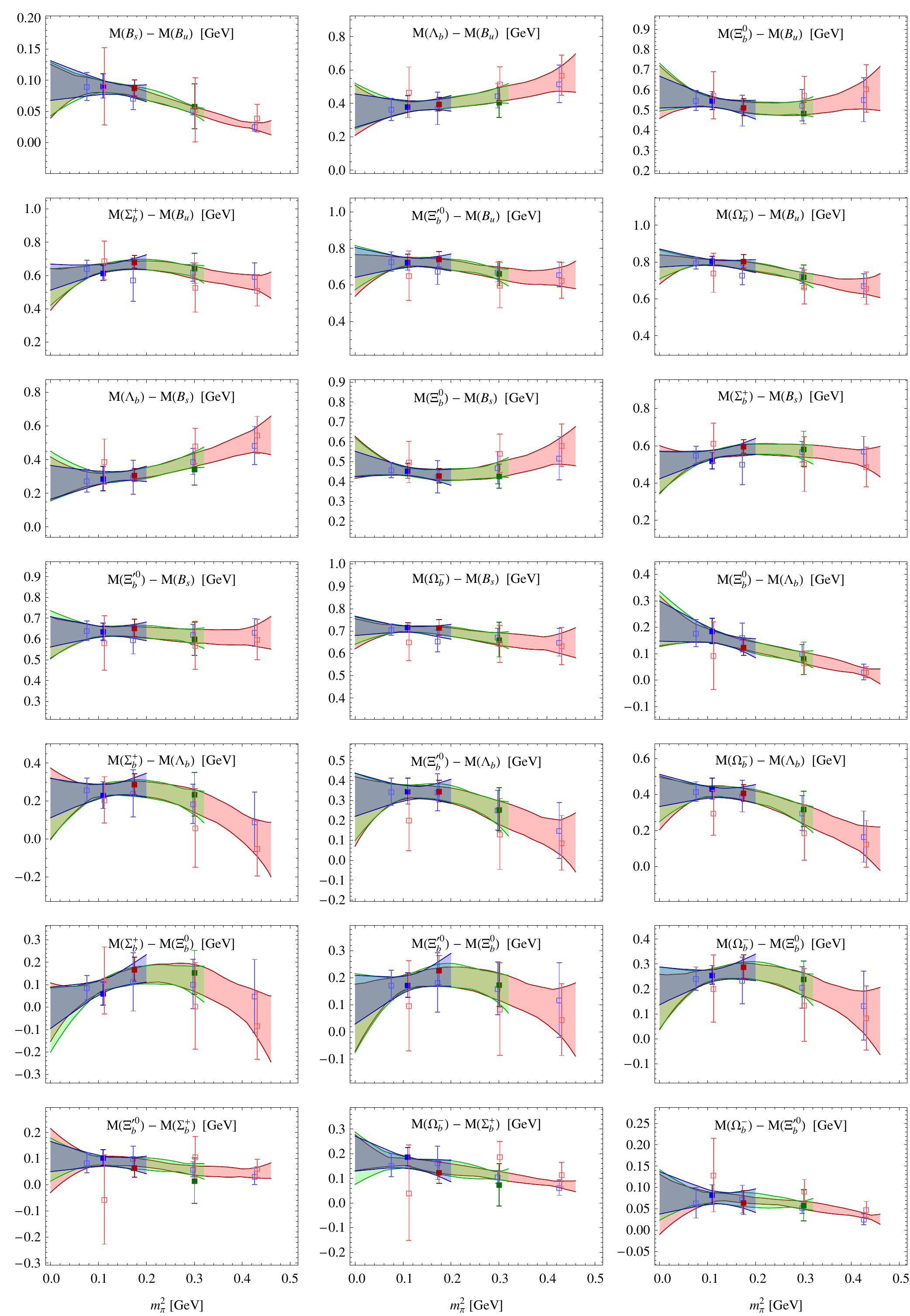}
    \caption{Polynomial extrapolations of the mass splittings
      calculated on the various ensembles are shown as a function of
      the pion mass. The blue, red and green
      points correspond to ensembles A, B and C respectively. Solid
      (open) symbols denote QCD (partially-quenched QCD)
      calculations. The uncertainties on the lattice data combine the
      statistical uncertainties and the various systematic
      uncertainties in quadrature. The blue, green and red shaded
      regions show the one standard deviation allowed
        region of linear, quadratic and cubic fits to the data
      within the ordinate extent of the corresponding region.}
  \label{fig:massdep}
\end{figure}

\begin{figure}[!t]
  \centering
  \includegraphics[width=0.93\columnwidth]{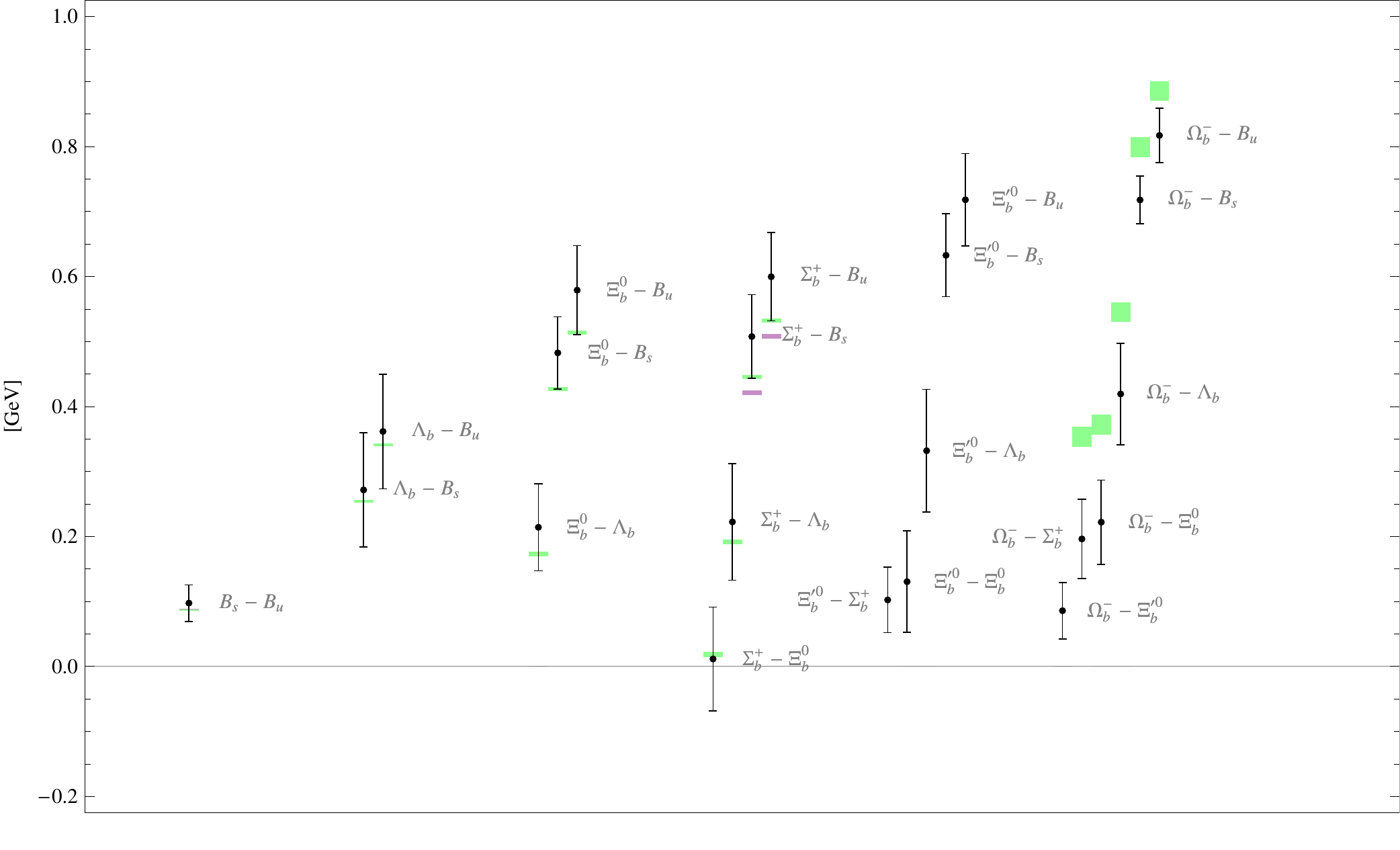}
  \caption{SU(3) mass splittings between bottom hadrons linearly extrapolated
    to the physical light quark masses. The green shaded regions correspond
    to the experimental determinations (combinations with both the
    $B_{u,s}^\ast$ and the $\Sigma_b^{\ast+}$ are shown in purple).}
  \label{fig:splittings}
\end{figure}

In Fig.~\ref{fig:massdep}, we show the results of the linear (blue),
quadratic (red) and cubic (green) fits to the data contained in the
ordinate extent of corresponding shaded region. The extracted
splittings at the physical light quark masses are presented in
Table~\ref{tab:splittings} for each extrapolation and compared to
experimental determinations where available. Differences between the
three forms of extrapolations are relatively small, and, as an example, the
linearly extrapolated mass differences are shown in
Fig~\ref{fig:splittings} along with the experimental determinations.
\begin{table}[h]
  \centering
 \begin{ruledtabular}
  \begin{tabular}{cccccc}
$h_1$ & $h_2$ & $M_{h_1}- M_{h_2}$ (lin.)  & $M_{h_1}- M_{h_2}$ (quad.) & $M_{h_1}- M_{h_2}$ (cub.) &
 $M_{h_1}- M_{h_2}$ (expt.)\\ \hline
$B_s$ & $B_u$ & 0.097(28) & 0.092(35) & 0.084(32) & 0.0867(11) \\
 $\Lambda_b$ & $B_u$ & 0.362(88) & 0.37(11) & 0.36(12) & 0.3407(20) \\
 $\Lambda_b$ & $B_s$ & 0.271(88) & 0.28(12) & 0.29(10) & 0.2540(22) \\
 $\Xi_b^0$ & $B_u$ & 0.579(69) & 0.595(94) & 0.588(98) & 0.5136(30) \\
 $\Xi_b^0$ & $B_s$ & 0.482(55) & 0.500(75) & 0.502(82) & 0.4269(32) \\
 $\Xi_b^0$ & $\Lambda_b$ & 0.214(67) & 0.211(86) & 0.212(76) & 0.1729(36) \\
 $\Sigma_b^+$ & $B_u$ & 0.600(68) & 0.564(89) & 0.556(93) & 0.5322(30) \\
 $\Sigma_b^+$ & $B_s$ & 0.508(64) & 0.484(88) & 0.491(96) & 0.4455(32) \\
 $\Sigma_b^+$ & $\Lambda_b$ & 0.222(90) & 0.19(13) & 0.21(14) & 0.1915(36) \\
 $\Sigma_b^+$ & $\Xi_b^0$ & 0.012(80) & -0.02(11) & -0.002(93) & 0.0186(42) \\
 $\Xi_b^{\prime 0}$ & $B_u$ & 0.718(71) & 0.698(98) & 0.677(90) & --- \\
 $\Xi_b^{\prime 0}$ & $B_s$ & 0.633(64) & 0.622(91) & 0.616(79) & --- \\
 $\Xi_b^{\prime 0}$ & $\Lambda_b$ & 0.332(94) & 0.30(13) & 0.27(13) & --- \\
 $\Xi_b^{\prime 0}$ & $\Xi_b^0$ & 0.131(78) & 0.10(11) & 0.094(95) & --- \\
 $\Xi_b^{\prime 0}$ & $\Sigma_b^+$ & 0.102(50) & 0.094(64) & 0.079(92) & --- \\
 $\Omega_b^-$ & $B_u$ & 0.817(42) & 0.794(61) & 0.776(58) & 0.886(15) \\
 $\Omega_b^-$ & $B_s$ & 0.718(37) & 0.705(48) & 0.696(51) & 0.799(15) \\
 $\Omega_b^-$ & $\Lambda_b$ & 0.419(78) & 0.385(98) & 0.37(11) & 0.545(15) \\
 $\Omega_b^-$ & $\Xi_b^0$ & 0.222(65) & 0.182(97) & 0.175(83) & 0.372(15) \\
 $\Omega_b^-$ & $\Sigma_b^+$ & 0.196(61) & 0.181(82) & 0.173(55) & 0.354(15) \\
 $\Omega_b^-$ & $\Xi_b^{\prime 0}$ & 0.086(43) & 0.084(47) & 0.076(55) & ---
  \end{tabular}
\end{ruledtabular}
\caption{Extracted static limit mass differences [GeV] obtained through 
  linear, quadratic or cubic extrapolations in $m_\pi^2$ to its
  physical value. The uncertainties in  the extrapolated values
  correspond to  one standard deviation allowed region of the space of
  fit parameters. 
  Experimental mass splittings \cite{Amsler:2008zz,Abazov:2008qm} are shown for
  comparison (for the mesons and the sextet baryons, we report the lighter of the 
  heavy quark spin multiplets). The  $\Xi_b^{\prime 0}$ has not been
  observed.}
  \label{tab:splittings}
\end{table}

\section{Conclusions and Outlook}
\label{sec:conclusions-outlook}

In this paper, we have presented a study of the mass splittings
between mesons and baryons containing one static quark using lattice
calculations based on the domain-wall fermion RBC/UKQCD ensembles. We have 
used domain wall quarks with a range of masses, the lightest set
corresponding to a pion mass of $\sim$275~MeV and at a single
lattice spacing, $a=0.114$~fm.  In general, the splittings we extract
agree well with experiment. Those involving the $\Omega_b^-$ tend to
be smaller than found experimentally as also found recently in the
lattice calculations of Refs.~\cite{Lewis:2008fu,Burch:2008qx}, but
the discrepancy is barely significant at our current level of precision
once the quark mass extrapolations are accounted for.

Our chiral extrapolations are somewhat {\it ad hoc} at present as the
axial couplings between $b$-hadrons and light pseudoscalar mesons that
appear in HH$\chi$PT are relatively unknown, making extrapolations
using the results of chiral perturbation theory problematic. Ongoing
calculations of these charges will allow us to control better the mass
difference extrapolations.  Additionally, the effects of the slightly
unphysical strange quark mass need to be accounted for.  Our
calculations are performed in the static limit of the $b$ quark and so
our extracted values for the splittings are subject to uncertainties
${\cal O}(\Lambda_{\rm QCD}/m_b)$ when compared to the experimental
bottom hadron spectrum. In the future, we intend to revisit these
calculations using either non-relativistic QCD or the Fermilab heavy quark
formalism. We also intend to extend our calculations to other
lattice spacings and volumes as they become available.

\acknowledgments{ We thank M.~J.~Savage and B.~C.~Tiburzi for useful
  conversations, and R.~Edwards and B.~Joo for help with, and the development
  of, the QDP++/Chroma programming environment~\cite{Edwards:2004sx}.  We are
  indebted to the RBC and UKQCD collaborations for use of their gauge
  configurations.  The work of WD was supported by the U.S.~Dept.~of Energy
  under Grant No.~DE-FG03-97ER41014. C-JDL was supported by the Taiwanese
  National Science Council via Grant No.~96-2112-M-009-020-MY3 and by
  CTS-North Taiwan. The computations for this work were performed at NERSC
  (Office of Science of the U.S. Department of Energy, No. DE-AC02-05CH11231),
  TeraGrid resources provided by the NCSA (thanks to the National Science
  Foundation), and the HPC facilities at National Chiao-Tung University. This
  work has also made use of resources provided by the Darwin Supercomputer
  of the University of Cambridge High Performance Computing Service, provided
  by Dell using Strategic Research Infrastructure Funding from the Higher
  Education Funding Council for England. We thank University of Cambridge,
  NCTS Taiwan and University of Washington for hospitality and travel support
  during the progress of this work.  }

%
%


\begin{thebibliography}{99}

\bibitem{Abazov:2008qm}
 V.~M.~Abazov {\it et al.}  [D0 Collaboration],
 arXiv:0808.4142 [hep-ex].

\bibitem{Barberio:2008fa}
  E.~Barberio {\it et al.}  [Heavy Flavor Averaging Group],
  arXiv:0808.1297 [hep-ex].




\bibitem{Bowler:1996ws}
 K.~C.~Bowler {\it et al.}  [UKQCD Collaboration],
 Phys.\ Rev.\  D {\bf 54}, 3619 (1996)
 [arXiv:hep-lat/9601022].

\bibitem{AliKhan:1999yb}
  A.~Ali Khan {\it et al.},
  Phys.\ Rev.\  D {\bf 62}, 054505 (2000)
  [arXiv:hep-lat/9912034].

\bibitem{Lewis:2001iz}
 R.~Lewis, N.~Mathur and R.~M.~Woloshyn,
 Phys.\ Rev.\  D {\bf 64}, 094509 (2001)
 [arXiv:hep-ph/0107037].

\bibitem{Mathur:2002ce}
 N.~Mathur, R.~Lewis and R.~M.~Woloshyn,
 Phys.\ Rev.\  D {\bf 66}, 014502 (2002)
 [arXiv:hep-ph/0203253].

\bibitem{Na:2007pv}
 H.~Na and S.~A.~Gottlieb,
 PoS {\bf LAT2007}, 124 (2007)
 [arXiv:0710.1422 [hep-lat]].

\bibitem{Lewis:2008fu}
 R.~Lewis and R.~M.~Woloshyn,
 arXiv:0806.4783 [hep-lat].

\bibitem{Burch:2008qx}
  T.~Burch, C.~Hagen, C.~B.~Lang, M.~Limmer and A.~Schafer,
  arXiv:0809.1103 [hep-lat].

\bibitem{Na:2008hz}
  H.~Na and S.~Gottlieb,
  arXiv:0812.1235 [hep-lat].

\bibitem{Manohar:2000dt}
 A.~V.~Manohar and M.~B.~Wise,
 Camb.\ Monogr.\ Part.\ Phys.\ Nucl.\ Phys.\ Cosmol.\  {\bf 10}, 1 (2000).

\bibitem{Antonio:2006px}
  D.~J.~Antonio {\it et al.}  [RBC and UKQCD Collaborations],
  Phys.\ Rev.\  D {\bf 75}, 114501 (2007).

\bibitem{RBCUKQCD}
C. Allton et al., Phys. Rev. D76 (2007) 014504; 
C. Allton et al., arXiv:0804.0473 [hep-lat].

\bibitem{Iwasaki:1984cj}
  Y.~Iwasaki and T.~Yoshie,
  Phys.\ Lett.\  B {\bf 143}, 449 (1984).

\bibitem{Iwasaki:1985we}
  Y.~Iwasaki,
  Nucl.\ Phys.\  B {\bf 258}, 141 (1985).


\bibitem{Kaplan:1992bt}
  D.~B.~Kaplan,
  Phys.\ Lett.\  B {\bf 288}, 342 (1992)
  [arXiv:hep-lat/9206013].

\bibitem{Shamir:1992im}
  Y.~Shamir,
  Phys.\ Lett.\  B {\bf 305}, 357 (1993)
  [arXiv:hep-lat/9212010].

\bibitem{Shamir:1993zy}
  Y.~Shamir,
  Nucl.\ Phys.\  B {\bf 406}, 90 (1993)
  [arXiv:hep-lat/9303005].

\bibitem{Furman:1994ky}
  V.~Furman and Y.~Shamir,
  Nucl.\ Phys.\  B {\bf 439}, 54 (1995)
  [arXiv:hep-lat/9405004].

\bibitem{Shamir:1998ww}
  Y.~Shamir,
  Phys.\ Rev.\  D {\bf 59}, 054506 (1999)
  [arXiv:hep-lat/9807012].


\bibitem{Teper:1987wt}
  M.~Teper,
  Phys.\ Lett.\  B {\bf 183}, 345 (1987).

\bibitem{Albanese:1987ds}
  M.~Albanese {\it et al.},
  Phys.\ Lett.\  B {\bf 192}, 163 (1987).


\bibitem{DellaMorte:2003mn}
  M.~Della Morte {\it et al.},
                  [ALPHA Collaboration],
  Phys.\ Lett.\  B {\bf 581}, 93 (2004).

\bibitem{Hasenfratz:2001hp}
 A.~Hasenfratz and F.~Knechtli,
 Phys.\ Rev.\  D {\bf 64}, 034504 (2001)
 [arXiv:hep-lat/0103029].

\bibitem{Amsler:2008zz}
 C.~Amsler {\it et al.}  [Particle Data Group],
 Phys.\ Lett.\  B {\bf 667}, 1 (2008).

\bibitem{Detmold:2008aaa}
  W.~Detmold \textit{et al.}, [NPLQCD Collaboration], in preparation.


\bibitem{Detmold:2008yn}
  W.~Detmold, K.~Orginos, M.~J.~Savage and A.~Walker-Loud,
  Phys.\ Rev.\  D {\bf 78}, 054514 (2008).

\bibitem{Prelovsek:2008rf}
  S.~Prelovsek and D.~Mohler,
  arXiv:0810.1759 [hep-lat].


\bibitem{Loktik:2006kz}
  O.~Loktik and T.~Izubuchi,
  Phys.\ Rev.\  D {\bf 75}, 034504 (2007)
  [arXiv:hep-lat/0612022].

\bibitem{Christ:2007cn}
  N.~H.~Christ, T.~T.~Dumitrescu, O.~Loktik and T.~Izubuchi,
  PoS {\bf LAT2007}, 351 (2007).

\bibitem{Martinelli:1998vt}
  G.~Martinelli and C.~T.~Sachrajda,
  Nucl.\ Phys.\  B {\bf 559}, 429 (1999)
  [arXiv:hep-lat/9812001].


\bibitem{Martinelli:1995vj}
 G.~Martinelli and C.~T.~Sachrajda,
 Phys.\ Lett.\  B {\bf 354}, 423 (1995)
 [arXiv:hep-ph/9502352].


\bibitem{Fleming:2004hs}
  G.~T.~Fleming,
  arXiv:hep-lat/0403023.

\bibitem{Wingate:2002fh}
 M.~Wingate {\it et al.},
 Phys.\ Rev.\  D {\bf 67}, 054505 (2003)
 [arXiv:hep-lat/0211014].



\bibitem{Lepage:2001ym}
  G.~P.~Lepage {\it et al.},
  Nucl.\ Phys.\ Proc.\ Suppl.\  {\bf 106}, 12 (2002)
  [arXiv:hep-lat/0110175].

\bibitem{Burdman:1992gh}
  G.~Burdman and J.~F.~Donoghue,
  Phys.\ Lett.\  B {\bf 280}, 287 (1992).
\bibitem{Wise:1992hn}
  M.~B.~Wise,
  Phys.\ Rev.\  D {\bf 45}, 2188 (1992).
\bibitem{Yan:1992gz}
  T.~M.~Yan {\it et al.}, 
  Phys.\ Rev.\  D {\bf 46}, 1148 (1992)
  [Erratum-ibid.\  D {\bf 55}, 5851 (1997)].

\bibitem{Cho:1992gg}
  P.~L.~Cho,
  Phys.\ Lett.\  B {\bf 285}, 145 (1992)
  [arXiv:hep-ph/9203225].
\bibitem{Cho:1992cf}
  P.~L.~Cho,
  Nucl.\ Phys.\  B {\bf 396}, 183 (1993)
  [Erratum-ibid.\  B {\bf 421}, 683 (1994)]
  [arXiv:hep-ph/9208244].
\bibitem{Savage:1995dw}
  M.~J.~Savage,
  Phys.\ Lett.\  B {\bf 359}, 189 (1995)
  [arXiv:hep-ph/9508268].
\bibitem{Chiladze:1997uq}
  G.~Chiladze,
  Phys.\ Rev.\  D {\bf 57}, 5586 (1998)
  [arXiv:hep-ph/9704426].

\bibitem{Tiburzi:2004kd}
  B.~C.~Tiburzi,
  Phys.\ Rev.\  D {\bf 71}, 034501 (2005)
  [arXiv:hep-lat/0410033].

  \bibitem{Edwards:2004sx} R.~G.~Edwards and B.~Joo,
Nucl.\ Phys.\ Proc.\ Suppl.\ {\bf 140} (2005) 832;
C. McClendon, "Optimized Lattice QCD Kernels for a Pentium 4 Cluster", 
Jlab preprint, JLAB-THY-01-29.

\end{thebibliography}
\end{document}